# Spin excitations in hole-overdoped iron-based superconductors

K. Horigane[1*], K. Kihou[2], K. Fujita[3], R. Kajimoto[4], K. Ikeuchi[5], S. Ji[6], J. Akimitsu[7], and C. H. Lee[2*]

[1]Graduate School of Natural Science and Technology, Okayama University, Okayama 700-8530, Japan. [2]National Institute of Advanced Industrial Science and Technology (AIST), Tsukuba, Ibaraki 305-8568, Japan. [3]Aoyama Gakuin University, Sagamihara, Kanagawa 229-8558, Japan. [4]J-PARC Center, Japan Atomic Energy Agency, Tokai, Ibaraki 319-1195, Japan. [5]CROSS, Tokai, Ibaraki 319-1106, Japan. [6]Max Plank POSTECH Center for Complex Phase Materials, Pohang University of Science and Technology, Pohang 37673, Republic of Korea. [7]Research Center of New Functional Materials for Energy Production, Storage, and Transport, Okayama University, Okayama 700-8530, Japan.

[*]E-mail:k-horigane@okayama-u.ac.jp; c.lee@aist.go.jp

**Understanding the overall features of magnetic excitation is essential for clarifying the mechanism of Cooper pair formation in iron-based superconductors. In particular, clarifying the relationship between magnetism and superconductivity is a central challenge because magnetism may play a key role in their exotic superconductivity. $BaFe_2As_2$ is one of ideal systems for such investigation because its superconductivity can be induced in several ways, allowing a comparative examination. Here we report a study on the spin fluctuations of the hole-overdoped iron-based superconductors $Ba_{1-x}K_xFe_2As_2$ (x = 0.5 and 1.0; $T_c$ = 36 K and 3.4 K, respectively) over the entire Brillouin zone using inelastic neutron scattering. We find that their spin spectra consist of spin wave and chimney-like dispersions. The chimney-like dispersion can be attributed to the itinerant character of magnetism. The band width of the spin wave-like dispersion is almost constant**

**from the non-doped to optimum-doped region, which is followed by a large reduction in the overdoped region. This suggests that the superconductivity is suppressed by the reduction of magnetic exchange couplings, indicating a strong relationship between magnetism and superconductivity in iron-based superconductors.**

**Introduction**

Spin-mediated superconductivity is one of the plausible models explaining the formation of Cooper pairs in iron-based superconductors. To investigate this hypothesis, magnetism has been intensively studied. The neutron scattering technique is a powerful method of examining spin fluctuations, because it can clarify both the energy and momentum dependences over the entire Brillouin zone. Revealing the overall spectrum of spin fluctuation is essential for understanding their magnetism and the mechanism of spin-mediated superconductivity.

The magnetism dependence of superconductivity should be demonstrated to prove that the superconductivity is due to spin fluctuations. $AFe_2As_2$ (A = Ba, Sr or Ca) is one of ideal systems for this purpose, because its superconductivity can be induced in several ways: electron doping [1,2], hole doping [3], chemical pressure [4] and external pressure [5]. Systematic and comparative studies among those samples can solve the problem.

The antiferromagnetic (AF) long-range ordering commonly observed in the parent compounds disappears and $T_c$ increases by applying pressure or upon doping in the underdoped region [5,6]. Apparently, AF long-range ordering competes against superconductivity. In the overdoped region, on the other hand, it is unclear why $T_c$ decreases and the superconductivity disappears with increasing doping. Inelastic neutron scattering (INS) studies on electron-doped $Ba(Fe,Ni)_2As_2$ have clarified that the spin fluctuations at high-energy region (>100 meV) are insensitive to doping [7] whereas those at low-energy region (<50 meV) disappear as $T_c$ decreases in the overdoped region [8]. It has been demonstrated that low-energy spin fluctuations are correlated with superconductivity. For hole-doped (Ba,K)Fe$_2$As$_2$, on the other hand, low-energy spin fluctuations remain even in heavily overdoped KFe$_2$As$_2$ below E = 20 meV [8,9]. Thus, why $T_c$ decreases upon hole doping in the hole-overdoped region remains as an unanswered question.

INS studies on the spin fluctuations of non-doped AFe$_2$As$_2$ (A = Ba, Sr or Ca) over the entire Brillouin zone have clarified that the spin dispersion can be well described by the $J_1$-$J_2$ model. However, it is unclear whether the model, which is based on a localized spin picture, is valid because carriers show an itinerant character. In fact, many calculations based on itinerant models have been proposed to explain the spin spectra [10,11,12]. Some models, for example based on the combination of density functional theory and dynamical mean-field theory, attempt to involve both itinerant and localized characters [13,14]. To establish a definitive model of magnetism, further examination of spin fluctuations is required.

Although systematic studies on spin fluctuations of electron-doped Ba(Fe,Ni)$_2$As$_2$ have been reported [7,15], the spin fluctuations of hole-overdoped samples over the entire Brillouin zone have not yet been established. In previous INS experiments on KFe$_2$As$_2$, spin fluctuations were observed up to E = 20meV, which is halfway to the zone boundary [8,9]. In optimum-doped (Ba,K)Fe$_2$As$_2$, a conflict between INS and resonant x-ray inelastic scattering (RIXS) has been found. The spin dispersion is robust upon doping according to the results of INS [8] whereas softening has been observed in RIXS experiments [16]. To solve this problem, further study of the hole doping dependence is essential. We, thus, report the overall spectra of spin fluctuations in overdoped Ba$_{1-x}$K$_x$Fe$_2$As$_2$ obtained by the INS technique.

**Results**

Figures 1(a)-(e) show the observed two-dimensional constant-energy images of spin excitations in the case of Ba$_{0.5}$K$_{0.5}$Fe$_2$As$_2$. Figures 1(f)-(j) show the results of calculations with peak positions derived from the Gaussian fitting of constant-energy spectra. We describe the (H, K) plane with orthorhombic notation, even though superconducting $Ba_{1-x}K_xFe_2As_2$ has a tetragonal crystal structure to facilitate comparison with non-doped $BaFe_2As_2$. Clear incommensurate peaks appeared around the (±1, 0) and (0, ±1) splitting along the longitudinal direction with a wave vector of (±2δ, 0) and (0, ±2δ), respectively, where δ = 0.06 at E = 13 meV. The (1±2δ, 0) position corresponds to [π(1±2δ,0)] in the ab plane and (0.5±δ, 0.5±δ) in tetragonal notation. As the energy increases, spin excitations start to split along the transverse direction corresponding to (±1±2δ, K) or (H, ±1±2δ) and reach the magnetic zone boundary with merging signals from next zone boundaries. In contrast, magnetic excitations along the longitudinal direction corresponding to (±1±2δ+H, 0) or (0, ±1±2δ+K) are strongly damped, consistent with previous observations for the $BaFe_2As_2$ system [17]. Figures 1(k, l) show the dispersion cuts along the transverse direction (1, K). A clear spin wave-like dispersion was observed up to E = 200 meV, where it reaches the zone boundary.

Figures 2(a)-(j) show two-dimensional constant-energy images of spin excitations and the results of calculations for x = 1. Well-defined incommensurate peaks with incommensurability larger than that for x = 0.5 are observed at E = 5meV. With increasing energy, spin excitations split along the transverse direction, similarly to the case of x = 0.5. The dispersion cut along the transverse direction (0.68, K) shows that the spin excitations reach the zone boundary around E = 80 meV (Fig. 2(k)). Nevertheless, magnetic signals exist even considerably above E = 80 meV with a vertical dispersion exhibiting a chimney-like structure (Figs. 2(l,m)). The energy-constant cuts along the K direction clearly show that the magnetic signals extend up to E ~ 200meV (Fig. 2(m)).

Next, we overview the overall spin dispersion of $Ba_{1-x}K_xFe_2As_2$. Figures 3(a)-(c) show the spin excitation dispersions for x = 0.5 and 1 at T = 6K derived from Gaussian fitting of the constant-energy spectra with those of the non- and underdoped samples. In x =

0.5, a spin wave dispersion is observed up to E = 200 meV, similarly to the cases of x = 0 and 0.33 (Fig. 3(a)). In x = 1, on the other hand, the dispersive spin excitations reach the zone boundary around E = 80 meV, which is considerably lower than the energy for x = 0.5 (Fig. 3(b)). Instead, a vertical dispersion with a chimney-like structure was observed from E = 80 meV up to 200 meV. In x = 0.5, signals of the chimney-like structure can also be found above E = 200 meV, but they are less clear than those in x = 1 (Fig. 1(m)).

Figure 3(d) shows the energy dependence of the dynamical magnetic susceptibility $\int\chi"(q,\omega)dq$ for x = 0.5 and 1 at T = 6 K. $\int\chi"(q, \omega)dq$ for x = 0, 0.33 ($T_c$ = 38.5 K) and $BaFe_{2-y}Ni_yAs_2$ (y = 0.18, $T_c$ = 8 K) reported in [7,8,15] are also depicted for comparison. It can be seen that $\int\chi"(q, \omega)dq$ for x = 0.5 exhibits essentially equivalent behavior to that for x = 0.33. Compared with the case of x = 0, on the other hand, the signals in the high-energy region are much lower for x = 0.5, while the peak energy remains around E = 150~200 meV. For x = 1, $\int\chi"(q, \omega)dq$ above E = 100 meV is further low, with the peak energy decreasing to around E = 30 meV. The large reduction in the high-energy spin fluctuations with hole doping results in suppression of the total fluctuating moment, which has been estimated to be $<m^2>$ = 1.45 and 0.65 $\mu_B^2$/Fe, for x = 0.5 and 1, respectively (Fig. 4). In contrast, $\int\chi"(q,\omega)dq$ in the low-energy region is almost independent of the doping level except for the sharp peak attributed to the spin resonance. Thus, the suppression of superconductivity in the hole-overdoped region cannot be due to a decrease in low-energy magnetic intensity as for electron-doped $Ba(Fe,Ni)_2As_2$ [7,8].

**Discussion**

The present observations demonstrate that the energy scale of the dispersive spin wave is robust upon hole doping up to x = 0.5, which is followed by a rapid decrease up to x = 1 (Fig. 4). The decrease appears to be related to the appearance of the incommensurate spin structure. In fact, the band width is robust in electron-doped

Ba(Fe,Co,Ni)$_2$As$_2$, which exhibits a commensurate spin structure except in the incommensurate AF state, which appears in a narrow doping range and has one-order smaller incommensurability than that of KFe$_2$As$_2$ [7,18]. The smaller band width of the spin wave leads to weaker effective magnetic exchange coupling $J$ according to the Heisenberg model. The results, thus, suggest that $J$ is correlated with the periodicity of spin fluctuations. The small value of $J$ in KFe$_2$As$_2$ is consistent with the fact that its electronic interaction strength $U$ is quite large [19,20].

The chimney-like structure can originate from particle-hole excitations, which define the itinerant character of spin fluctuations. Note that the chimney-like structure resembles the spin excitations in the itinerant AF metals Cr [21], Cr$_{0.95}$V$_{0.05}$ [22] and Mn$_{2.8}$Fe$_{0.2}$Si [23]. The present results show that the band width decreases and the chimney-like dispersion appears with hole doping. This is qualitatively consistent with DFT + DMFT calculations [13,14], which also supports the origin of the chimney-like dispersion to be particle-hole excitations.

The present observation of a large reduction in high-energy spin fluctuations upon hole doping is in contrast to the case of electron-doped Ba(Fe,Ni)$_2$As$_2$, where high-energy spin fluctuations are independent of doping [7]. Because hole-doped (Ba,K)Fe$_2$As$_2$ exhibits a higher maximum $T_c$ of 38 K than electron-doped Ba(Fe,Ni)$_2$As$_2$ ($T_c$ = 20 K) [2] even though (Ba,K)Fe$_2$As$_2$ exhibits weaker spin fluctuations in the high-energy region, this reduction of the high-energy spin fluctuations does not appear to suppress $T_c$. The suppression of $T_c$ in hole-overdoped (Ba,K)Fe$_2$As$_2$ can rather be attributed to the reduction of $J$, which remains almost constant from non-doped to optimum-doped region and followed by rapid reduction in the overdoped region. The $J$ dependence of the superconductivity has also been suggested in studies on spin resonance [24-26]. Stronger magnetic correlation leads to a larger energy split between the resonance and the superconducting gap energy. In fact, the resonance energy in overdoped (Ba,K)Fe$_2$As$_2$ approaches the superconducting gap energy with doping up to x = 0.77 [24], which can result from the reduction of $J$. These results lead to the conclusion that there is a strong relationship between magnetism and superconductivity in $Ba_{1-x}K_xFe_2As_2$.

**Method**

Single crystals of $Ba_{0.5}K_{0.5}Fe_2As_2$ ($T_c$ = 36 K) and $KFe_2As_2$ ($T_c$ = 3.4 K) were grown by a KAs self-flux method [27,28]. The magnetic susceptibility was measured by a SQUID from Quantum Design.

The INS measurement was performed using the Fermi chopper spectrometer 4SEASONS in J-PARC[29]. We co-aligned 160 and 300 pieces of single crystal with x = 0.5 (~5 g) and x = 1 (~5 g), respectively. We employed the multi-$E_i$ method [30] with incident neutron energies of $E_i$ = 31, 65, 110, 202, 409, 720 meV for $Ba_{0.5}K_{0.5}Fe_2As_2$ and $E_i$ = 30, 75, 149, 423 meV for $KFe_2As_2$. The incident beam was parallel to the c-axis. We converted signal intensities into absolute units using a vanadium standard. The data were processed by the "Utsusemi" visualization software developed at J-PARC [31]. Throughout this letter, wave vectors are specified in the orthorhombic reciprocal lattice.

[22] S. M. Hayden et al., Strongly enhanced magnetic excitations near the quantum critical point of $Cr_{1-x}V_x$ and why strong exchange enhancement need no imply heavy fermion behavior. Phys. Rev. Lett. **84**, 999 (2000).

[23] S. Tomiyoshi et al., Magnetic excitations in the itinerant antiferromagnets $Mn_3Si$ and Fe-doped $Mn_3Si$. Phys. Rev. B **36**, 2181 (1987).

[24] C. H. Lee et al., Suppression of spin-exciton state in hole overdoped iron-based superconductors. Sci. Rep. **6**, 23424 (2016).

[25] C. H. Lee et al., Universality of the dispersive spin-resonance mode in superconducting $BaFe_2As_2$. Phys. Rev. Lett. **111**, 167002 (2013).

[26] D. K. Pratt et al., Dispersion of the superconducting spin resonance in underdoped and antiferromagnetic $BaFe_2As_2$. Phys. Rev. B **81**, 140510(R) (2010).

[27] K. Kihou et al., Single crystal growth and characterization of the iron-based superconductor $KFe_2As_2$ synthesized by KAs flux method. J. Phys. Soc. Jpn. **79**, 124713 (2010).

[28] K. Kihou et al., Single-crystal growth of $Ba_{1-x}K_xFe_2As_2$ by KAs self-flux method. J. Phys. Soc. Jpn. **85**, 034718 (2016).

[29] R. Kajimoto et al., The Fermi chopper spectrometer 4SEASONS at J-PARC. J. Phys. Soc. Jpn. **80**, SB025 (2011).

[30] M. Nakamura et al., First demonstration of novel method for inelastic neutron scattering measurement utilizing multiple incident energies. J. Phys. Soc. Jpn. **78**, 093002 (2009).

[31] Y. Inamura, T. Nakatani, J. Suzuki and T. Otomo, Development status of software "Utsusemi" for chopper spectrometers at MLF, J-PARC. J. Phys. Soc. Jpn. **82**, SA031 (2013).

<>
**Acknowledgments**

We would like to acknowledge H. Hiraka, T. Fukuda, S. Onari, T. Tohyama and K. Yamada for valuable discussions. The neutron experiment at the Materials and Life Science Experimental Facility of J-PARC was performed under user programs (2012B0075 and 2013B0061). This work was supported by Grants-in-Aid for Scientific Research B (24340090, 25287081) and for Young Scientists B (16K17750) from Japan Society for the Promotion of Science.

**Author contributions**

K.H., C.H.L., K.F., R.K., K.I., and S.J. conducted the inelastic neutron scattering measurements and analyzed the data. K.H., K.K. and K.F. synthesized and characterized the single crystals. C.H.L., and J.A. designed and coordinated the experiment. All authors contributed to and discussed the manuscript.

**Additional Information**

Competing financial interests: The authors declare no competing financial interests.

Fig.1 Constant-energy images of spin excitations of $Ba_{0.5}K_{0.5}Fe_2As_2$ at T = 6K at (a) E = 13±0.5meV obtained with incident neutron energy of $E_i$ = 31meV; (b) E = 40±1meV with $E_i$ = 65meV; (c) E = 80±3meV with $E_i$ = 202meV; (d) E = 110±5meV with $E_i$ = 202meV; (e) E = 200±10meV with $E_i$ = 409meV. The color bars represent the vanadium-normalized absolute intensity ($d^2\sigma/(d\Omega dE)$) in units of mbarn sr$^{-1}$ meV$^{-1}$ f.u.$^{-1}$. (f)-(j) Simulated pictures of net intensity derived from obtained fitting parameters (peak intensity, peak position, FWHM and background) corrected by the magnetic form factor. The dashed box indicates the AF zone boundaries. (k), (l) Dispersion cuts of $Ba_{0.5}K_{0.5}Fe_2As_2$ with $E_i$ = 202 and 409 meV, respectively, along the K direction. The solid lines are a guide to the eye. (m) Constant-energy cuts of spin excitations along the K direction as a function of energy. The blue arrows represent peak positions of magnetic signals. Peaks observed on both sides without the blue arrows are magnetic signals originating from another magnetic Brillouin zone. Solid and dotted lines are the results of Gaussian fitting and the magnetic zone boundaries, respectively.

Fig.2 Constant-energy images of spin excitations of $KFe_2As_2$ at T = 6K at (a) E = 5±0.5meV obtained with $E_i$ = 30meV; (b) E = 15±0.5meV with $E_i$ = 30meV; (c) E = 38±1meV with $E_i$ = 75meV; (d) E = 76±2meV with $E_i$ = 149meV; (e) E = 160±20meV with $E_i$ = 423meV. (f)-(j) Simulated pictures of net intensity derived from obtained fitting parameters (peak intensity, peak position, FWHM and background) corrected by the magnetic form factor. The dashed box indicates the AF zone boundaries. (k) Dispersion cut of $KFe_2As_2$ with $E_i$ = 149 meV along *K* direction. The solid lines are guide to the eye. The vertical white dashed lines depict the magnetic zone boundaries. (l) High-energy spin excitation of $KFe_2As_2$ with $E_i$ = 423 meV. (m) Variation in the scattered intensity along the K direction for different excitation energies. The blue arrows represent peak positions of magnetic signals. Peaks observed on both sides without the blue arrows are magnetic signals originating from another magnetic Brillouin zone. Solid and dotted lines depict the results of Gaussian fitting and the magnetic zone boundaries, respectively.

Fig.3 Dispersions of spin excitations for $Ba_{0.5}K_{0.5}Fe_2As_2$ and $KFe_2As_2$ along transverse direction. (a)-(c) Spin excitation dispersions of $Ba_{0.5}K_{0.5}Fe_2As_2$ (red) and $KFe_2As_2$ (blue) at T = 6K plotted as filled circles. The horizontal and vertical lengths of the green rectangles represent the full width at half maximum of the Gaussian fits of the constant-energy spectrum and the energy range in the fitting, respectively. Scan directions are depicted by the arrows in the insets. The dashed and dotted lines in (c) are dispersion of $BaFe_2As_2$ at T = 7K (Ref.17) and $Ba_{0.67}K_{0.33}Fe_2As_2$ at T = 9K (Ref.8), respectively. The vertical dashed lines depict the magnetic zone boundaries for $Ba_{0.5}K_{0.5}Fe_2As_2$. (d) Energy dependence of $\int\chi''(q,\omega)dq$ for $Ba_{0.5}K_{0.5}Fe_2As_2$ (filled red circles) and $KFe_2As_2$ (filled blue circles) at T = 6K. The solid lines are a guide to the eye. The black dashed lines depict $\int\chi''(q,\omega)dq$ for $BaFe_2As_2$ at T = 5K (Ref.15), $Ba_{0.67}K_{0.33}Fe_2As_2$ at T = 9K (Ref.8) and $BaFe_{1.82}Ni_{0.18}As_2$ (Ref.7) at T=5K, respectively.

Fig.4 Phase diagram of $Ba_{1-x}K_xFe_2As_2$. Symbols denote $T_c$ (open circles, Ref.28), The Néel temperature $T_N$ (open squares, Ref.6), the total fluctuating magnetic moment $<m^2>$ (closed circles), band width of spin wave $W$ (closed squares), and the incommensurability $\delta$ (closed diamonds, Ref.24).

**Figure 1**

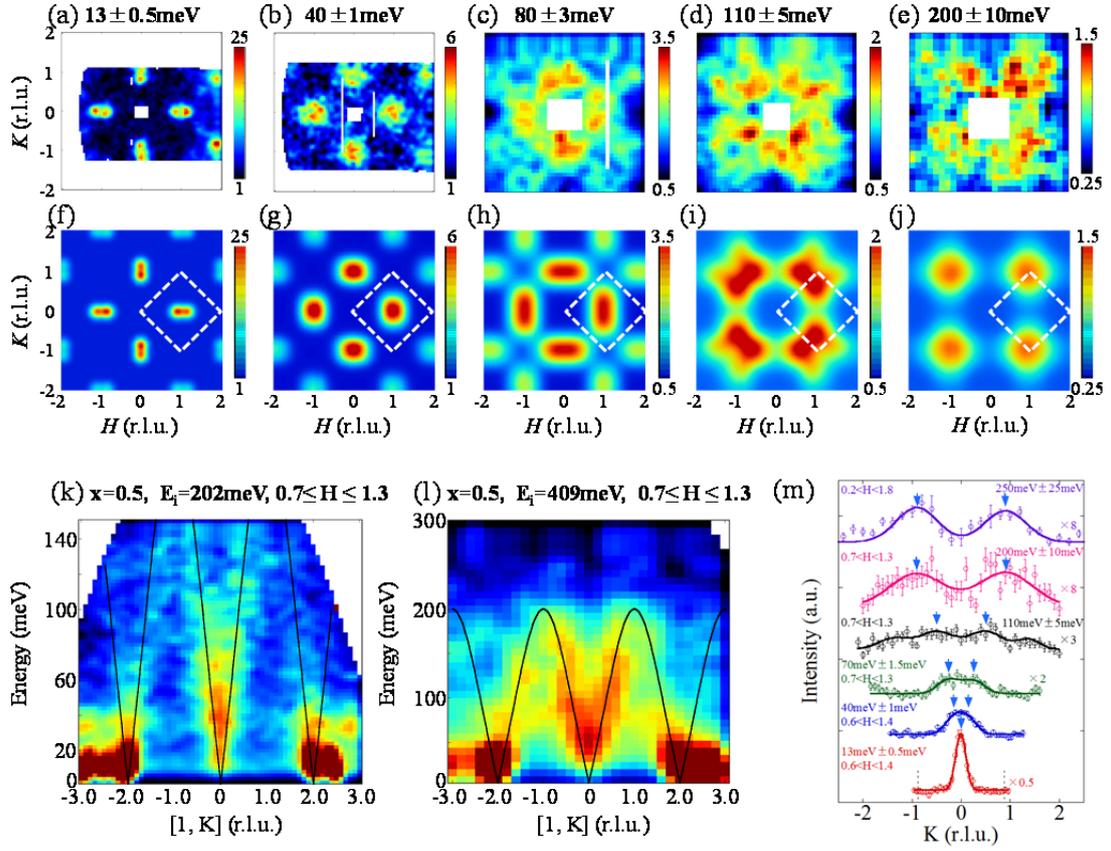

**Figure 2**

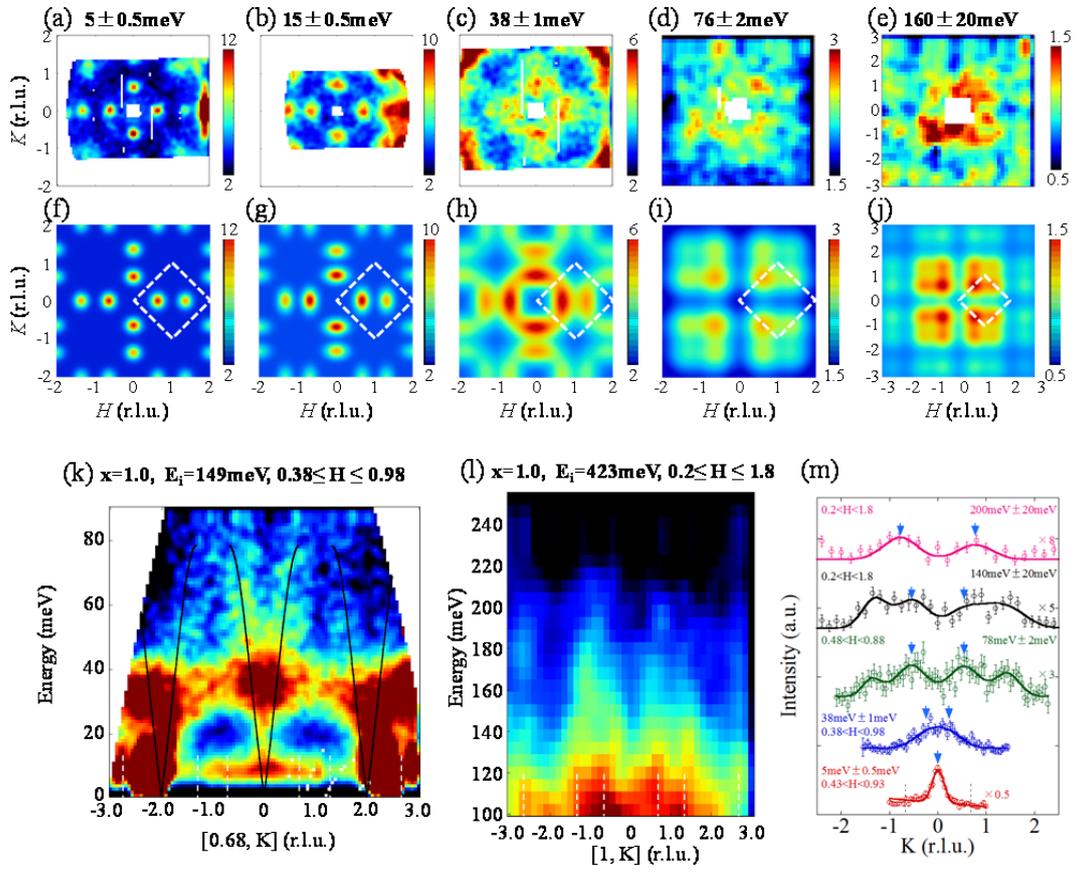

**Figure 3**

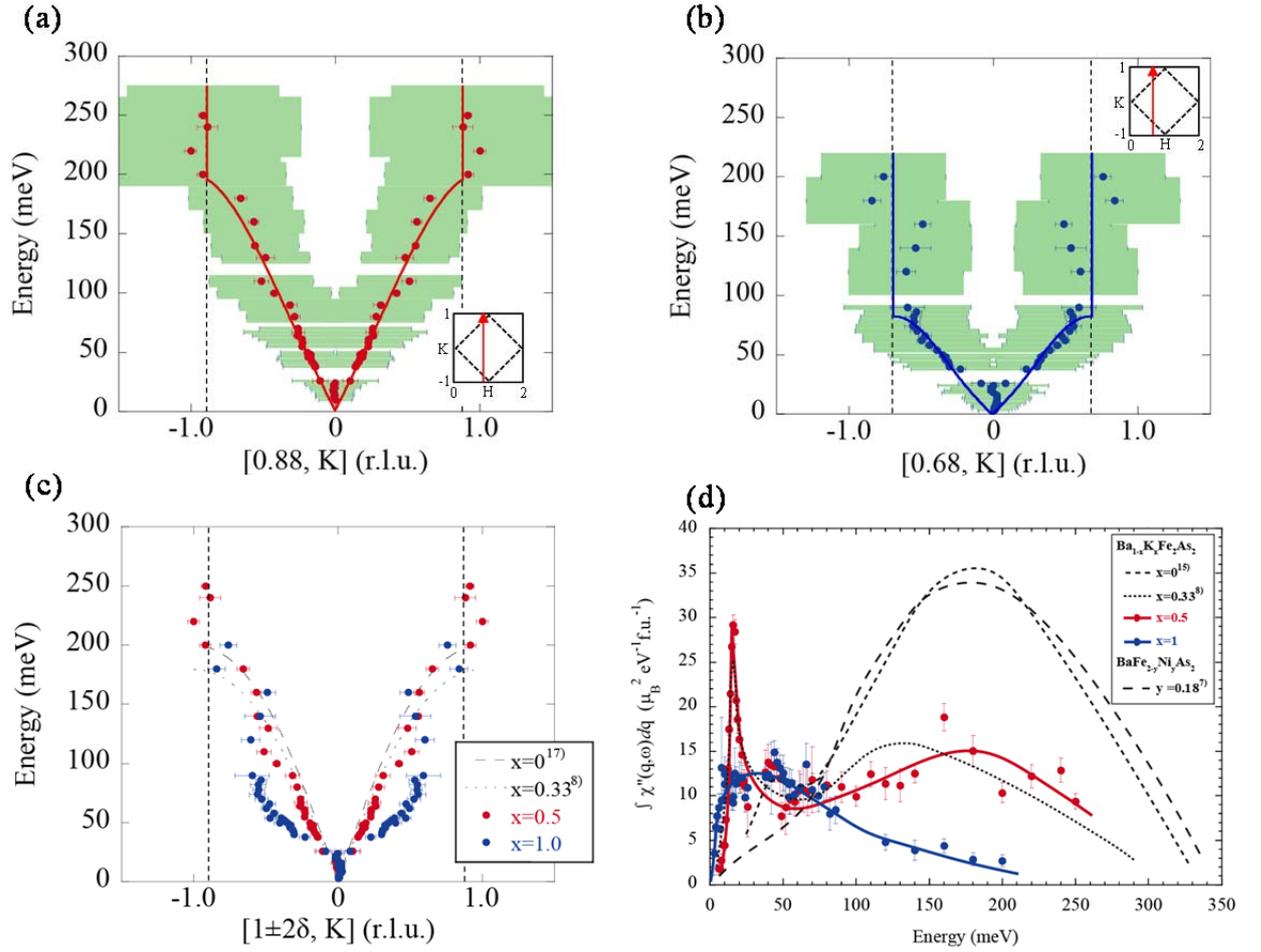

**Figure 4**

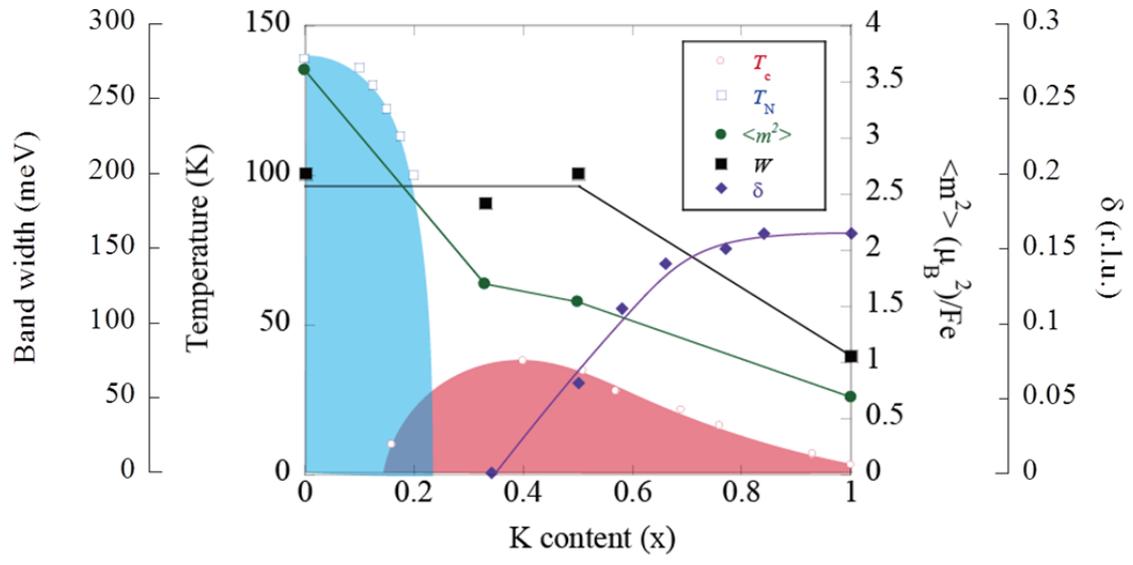